\documentclass{elsart}

\usepackage{natbib}

\usepackage{amssymb}

\usepackage{graphicx}
\newcommand{\ltsima}{$\; \buildrel < \over \sim \;$}
\newcommand{\simlt}{\lower.5ex\hbox{\ltsima}}
\newcommand{\gtsima}{$\; \buildrel > \over \sim \;$}
\newcommand{\simgt}{\lower.5ex\hbox{\gtsima}}

\newcommand{\cgs}{ ${\rm erg~cm}^{-2}~{\rm s}^{-1}$} 
\newcommand{\lum}{\rm erg s$^{-1}$}
\newcommand{\pn}{\par\noindent}

\def\lesssim{\mathrel{\hbox{\rlap{\hbox{\lower4pt\hbox{$\sim$}}}\hbox{$<$}}}}
\def\gtrsim{\mathrel{\hbox{\rlap{\hbox{\lower4pt\hbox{$\sim$}}}\hbox{$>$}}}}

\def\arcsec{\hbox{$^{\prime\prime}$}}

\def\ab1450{$AB_{1450(1+z)}$}

\def\xray{\hbox{X-ray}}

\def\chandra{{\it Chandra\/}}

\def\heao1{{\it HEAO-1\/}}

\def\xmm{{XMM-{\it Newton\/}}}

\def\mnras{Mon. Not. Roy. Astron. Soc.}
\def\apj{Astroph. Jou.}
\def\apjss{Astroph. Jou. Suppl. Ser.}
\def\aj{Astron. Jou.}
\def\aap{Astron \& Astroph.}

\def\na{New Astronomy}

\def\an{Astron. Nach.}

\begin{document}

\begin{frontmatter}



\title{On the Exotic Hard X-ray Source Populations in the Hellas2XMM survey}


\author{C. Vignali}
\ead{cristian.vignali@unibo.it}
\address{Dipartimento di Astronomia, Via Ranzani 1, 40127 Bologna, Italy}
\author{M. Mignoli}
\ead{marco.mignoli@bo.astro.it}
\address{INAF--Osservatorio Astronomico di Bologna, Via Ranzani 1, 
40127 Bologna, Italy}
\author{on behalf of the Hellas2XMM Collaboration\thanksref{h2xmm}}
\thanks[h2xmm]{A. Baldi, M. Brusa, N. Carangelo, P. Ciliegi, 
F. Civano, F. Cocchia, A. Comastri, F. Fiore, F. La Franca, R. Maiolino, 
G. Matt, S. Molendi, G.C. Perola, L. Pozzetti, S. Puccetti.}

\begin{abstract}

Recent hard \xray\ surveys have proven to be effective in discovering 
large numbers of \xray\ sources that, despite the likely association 
with active nuclei, appear to be characterized by ``peculiar'' properties. 
Among these ``exotic'' source populations, we will focus on the nature of 
two classes of hard \xray\ sources: those characterized by high 
X-ray-to-optical flux ratios -- a fraction of these are associated with 
the rather elusive Type~2 quasars -- and the \xray\ bright optically 
normal galaxies, also known as XBONGs. 

\end{abstract}


\end{frontmatter}

\section{Introduction}
\label{sect_intro}
One of the most interesting results of recent \xray\ surveys with \chandra\ 
and \xmm\ consists in the discovery of a large number of 
sources with ``peculiar'' properties, i.e., not straightforwardly associated 
with Active Galactic Nuclei (AGN). In particular, two classes of sources 
will be discussed in the following: (a) sources with high X-ray-to-optical 
flux ratios (X/O) and (b) \xray\ bright optically normal galaxies (XBONGs). 

\subsection{High X-ray-to-optical flux ratio sources}
\label{xo_subsec}
\begin{figure}
\centering
\includegraphics[angle=0,width=76mm]{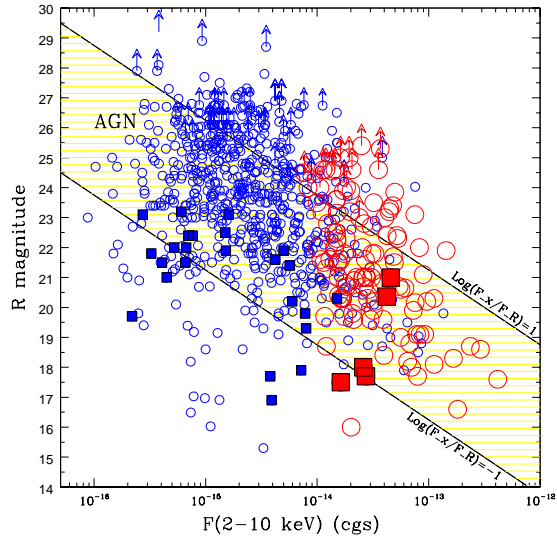}
\caption{2--10~keV flux vs. $R$-band magnitude for sources detected 
by recent \xray\ surveys 
[large symbols at bright \xray\ fluxes: Hellas2XMM results;  
small symbols at faint \xray\ fluxes: 
\chandra\ Deep Field-North and South \cite{bar03,szo04}, 
Lockman Hole \cite{mai02}, ELAIS \cite{man03}]. 
The squares indicate the sources classified as XBONGs. 
``Normal'' AGN typically populate the shaded region 
(within $\log F_{\rm X}/F_{\rm R}=\log (X/O)\pm{1}$).}
\label{fig1}
\end{figure}

In the \hbox{2--10~keV} flux vs. $R$-band magnitude plot (Fig.~\ref{fig1}), 
the high X/O sources populate the region above the dashed area 
(where AGN typically lie), being characterized by X/O$\ge$10. 
The fraction of this source population among \xray\ selected samples 
is $\approx$~20\% and constant over a broad range of \xray\ fluxes 
[e.g., \cite{fio03,eck04}] . 
From Fig.~\ref{fig1} it appears evident that a significant fraction of 
these high X/O sources (typically those at $R>25$) cannot be 
spectroscopically identified even with 10-m class telescopes. 
We took advantage of the relatively bright \xray\ flux limit of the 
Hellas2XMM survey [$F_{2-10~keV}\approx$~10$^{-14}$~\cgs; 
see \cite{bal02} for details] and, therefore, 
of the bright magnitudes of most of the optical counterparts, 
to shed light on the nature of these sources. 
We proceed through a two-steps method: firstly, we used FORS1 at the 
Very Large Telescope (VLT) to identify through spectroscopy the most 
luminous ($R<24$) members of this population. Of the whole sample of 
28 high X/O sources in the Hellas2XMM 1dF, 
13 were targeted \cite{fio03}. 
Eight sources, at $z\approx$~1--2, match the Type~2 quasar optical 
definition (i.e., luminous sources with high-ionization, 
narrow emission lines). 
Four sources are broad-line (Type~1) AGN, while 
the remaining object is a moderate-luminosity emission-line galaxy. 
\begin{figure}
\centering
\includegraphics[angle=-90,width=0.49\textwidth]{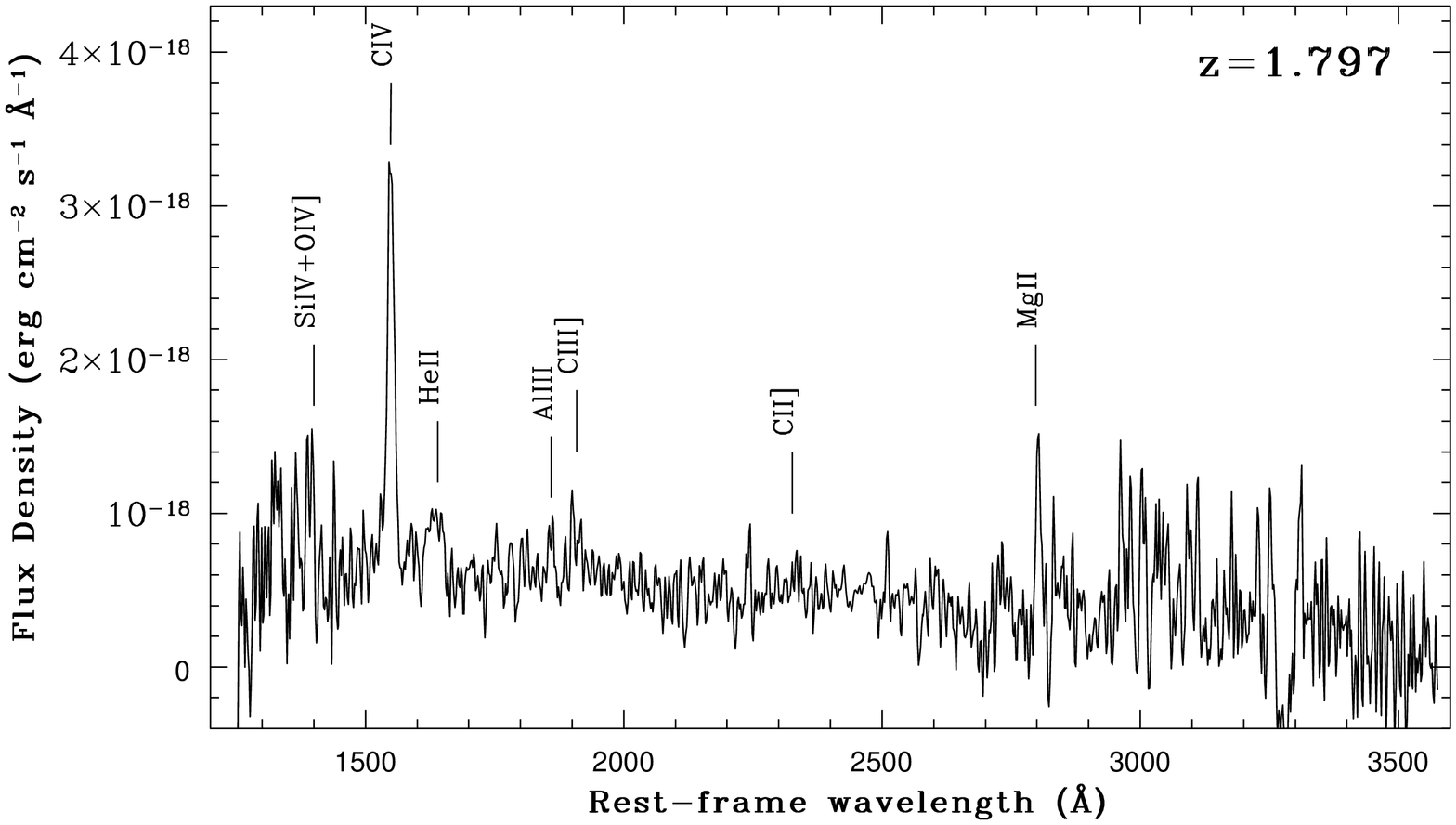}
\includegraphics[angle=-90,width=0.49\textwidth]{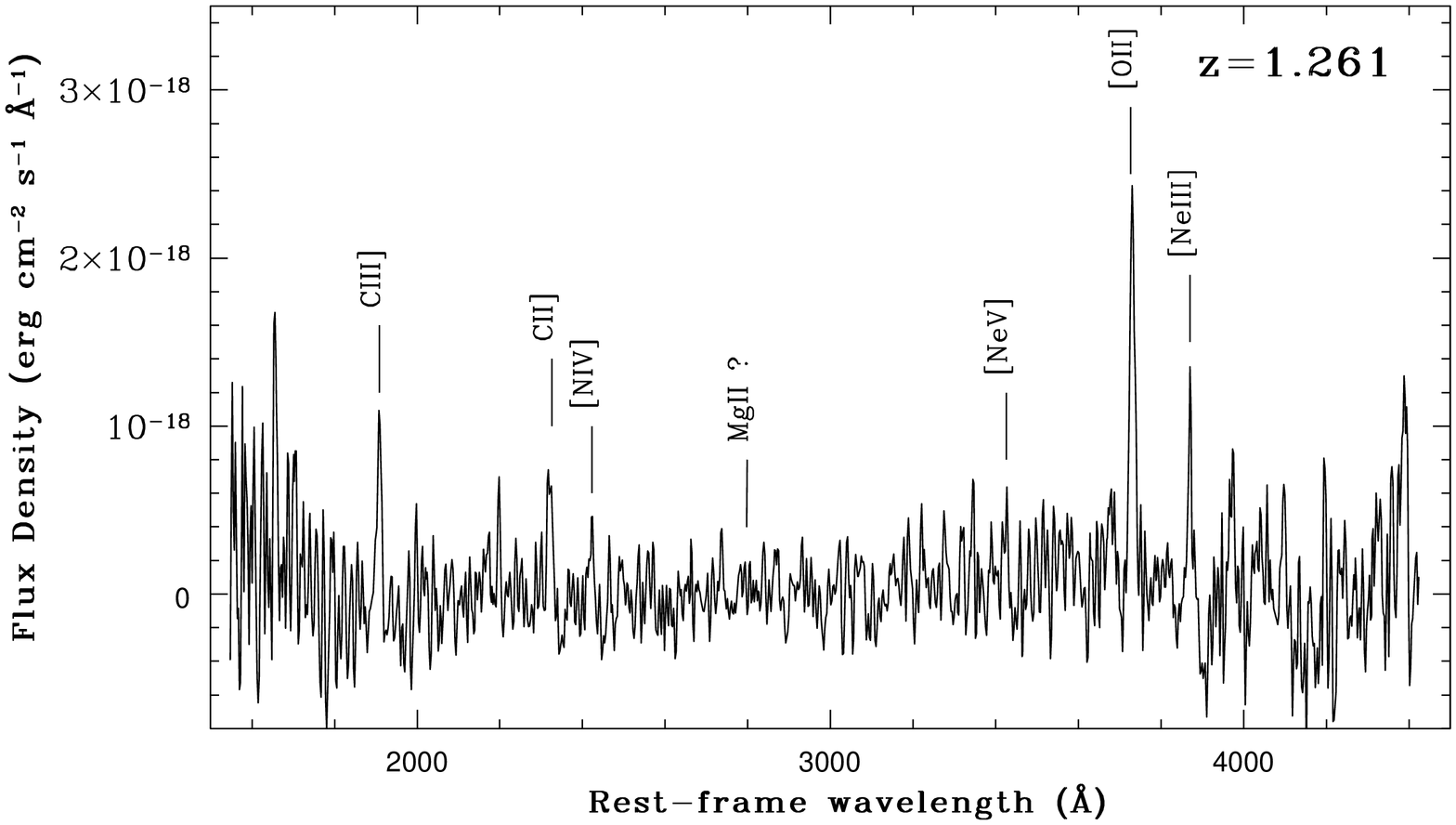}
\caption{Optical spectra of two Type~2 quasars discovered by the 
Hellas2XMM survey: 
source \#43 in PKS~0537$-$286 field (left panel) and 
source \#13 in Mrk~509 field (right panel). 
The principal emission lines are indicated.}
\label{fig2}
\end{figure}
From the \xray\ perspective, these objects match also the most widely 
accepted \xray\ definition of Type~2 quasars, i.e., they are 
luminous \hbox{($L_{2-10~keV}\gtrsim10^{44}$~\lum)} 
and absorbed [\hbox{$N_{\rm H}\gtrsim10^{22}$~cm$^{-2}$}; \cite{per04}]. 
In Fig.~\ref{fig2} two examples of Type~2 quasars from the Hellas2XMM 
surveys are shown. All the emission lines have FWHM$<$1,500~km~s$^{-1}$. 
Further information on the broad-band properties of these objects 
will be obtained with the planned sub-mm and near-infrared observations. 
Overall, although recent \xray\ [e.g., \cite{szo04,nor02,ste02}]
and optical surveys [\cite{zak03}; see also \cite{vig04} and 
references therein] have proven to be effective in the detection of many 
Type~2 quasars, much work needs to be done at the faintest optical 
magnitudes. The second step of our approach addresses this point. 
Eleven faint ($R\gtrsim24.5$) high X/O sources were observed with ISAAC at the 
VLT in the $K_{\rm S}$ band. Bright ($K_{\rm S}<19$) counterparts for 10 
sources were found. 
All the counterparts have extremely red colors ($R-K>5.0$); in particular, 
7 sources are associated with massive bulge-dominated galaxies and have 
estimated redshifts in the range \hbox{0.9--1.4} [see \cite{mig04} for 
details]. Since these objects are highly luminous and absorbed in the 
\xray\ band, they show {\em how 
effective the hard \hbox{X-rays} are in tracing obscured quasar 
accretion in massive ellipticals at \hbox{$z\approx$~1--2}}. 
Two of the remaining counterparts are point-like, 
possibly AGN at \hbox{$z\approx$~1.6--2.4}, and only one object has 
a disk morphology (at $z\approx$~0.8). 

\subsection{What's the nature of XBONGs?}
\label{xbong_subsec}
\begin{figure}
\centering
\includegraphics[angle=-90,width=0.49\textwidth,bb=60 20 575 720,clip]{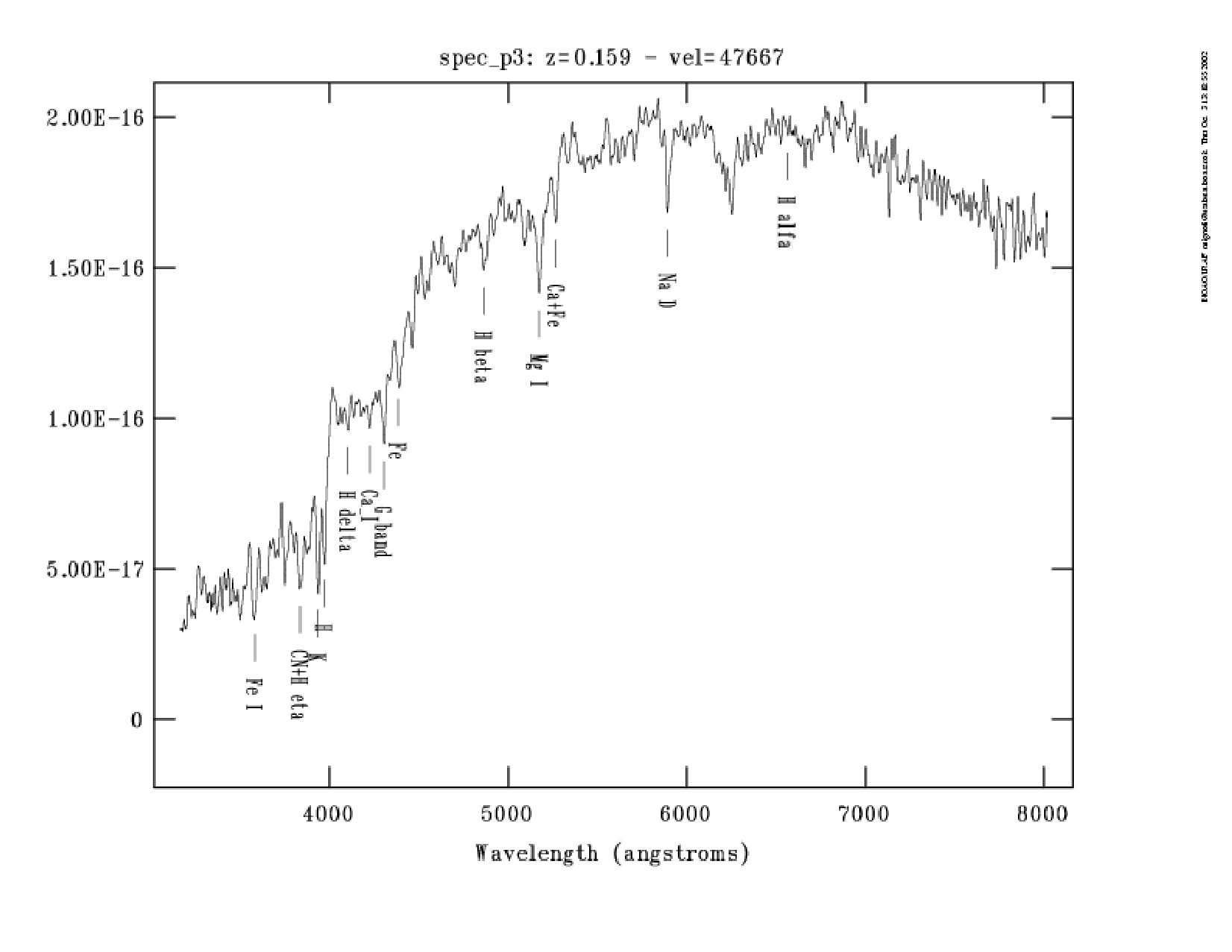}
\hfill \ 
\includegraphics[angle=-90,width=0.49\textwidth,bb=60 20 575 720,clip]{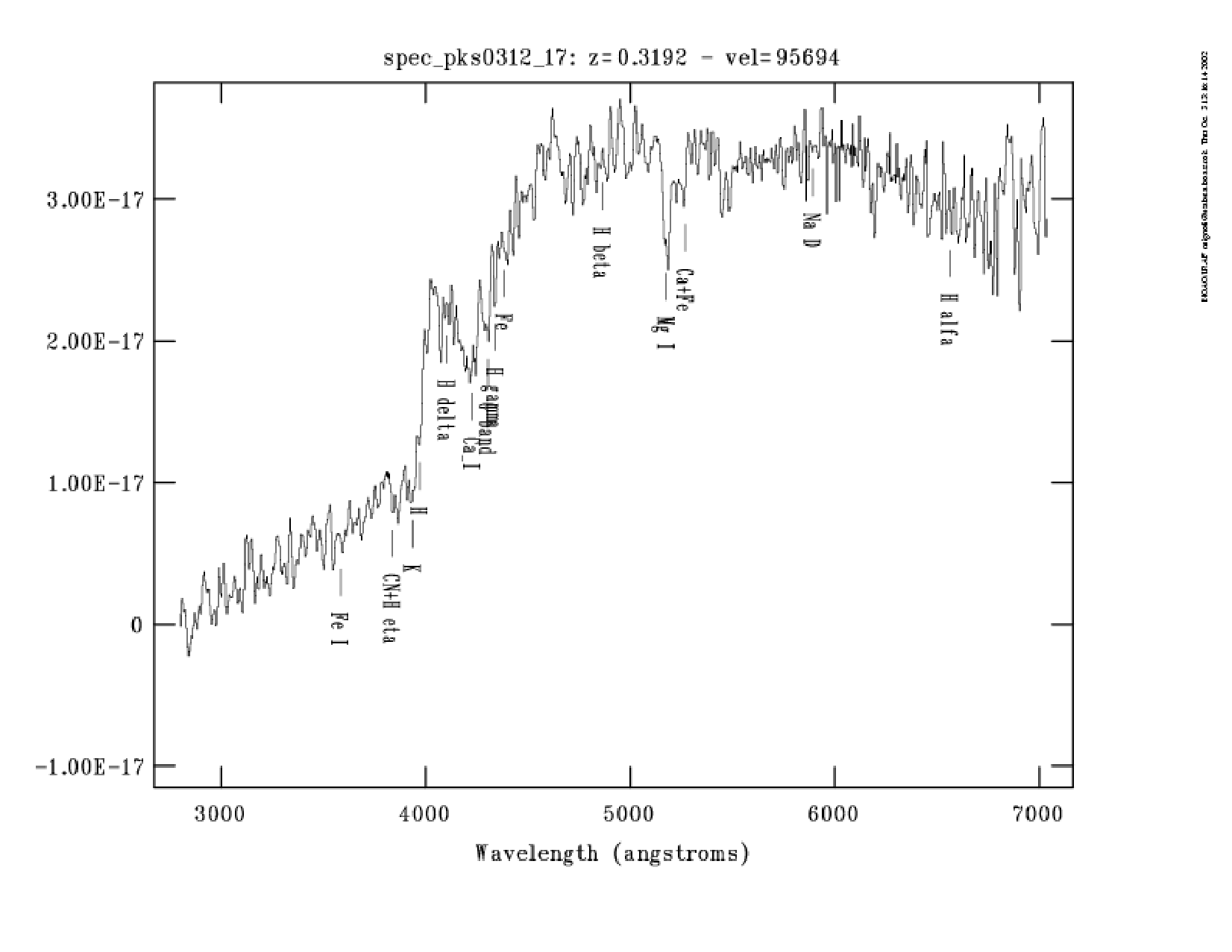}
\pn
\includegraphics[angle=-90,width=0.49\textwidth,bb=60 20 575 720,clip]{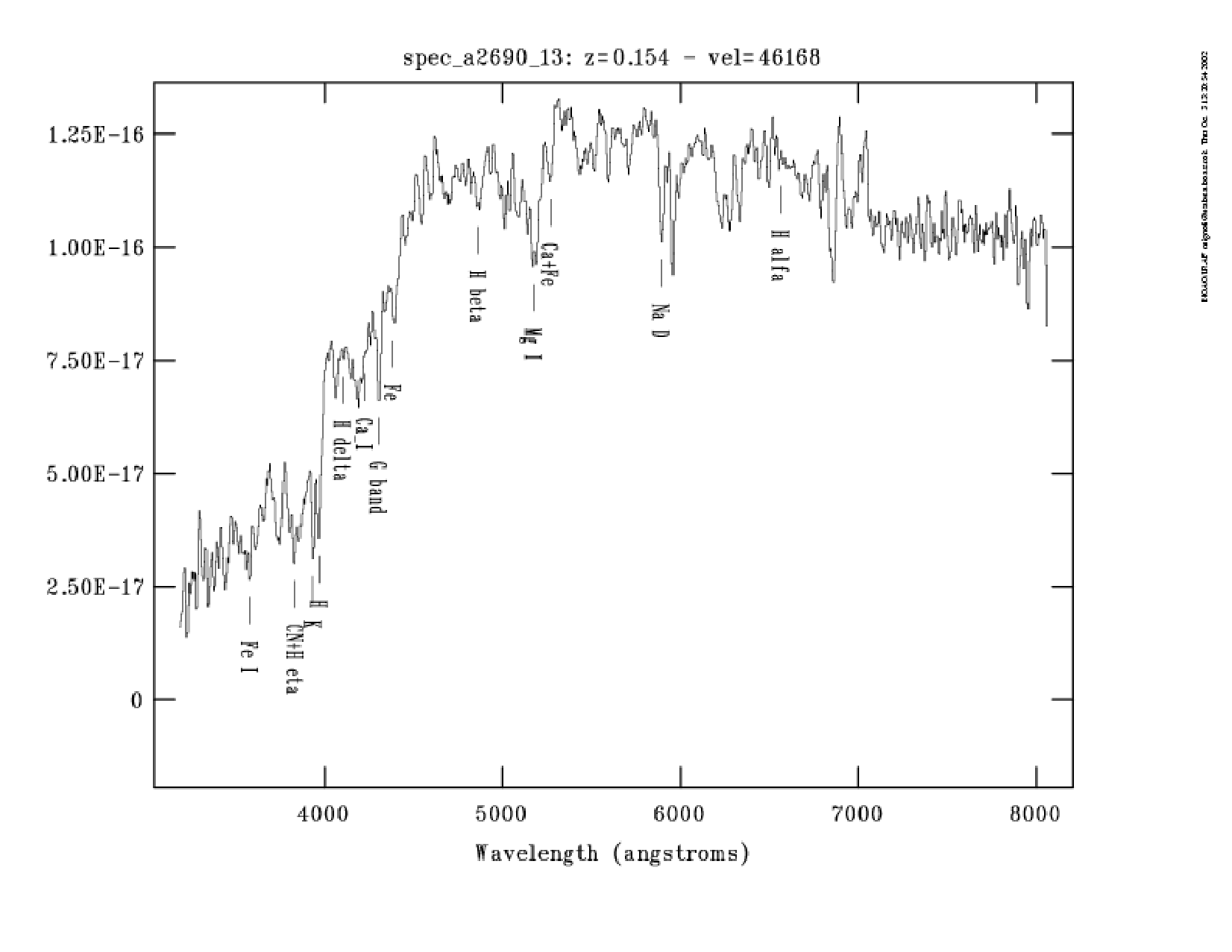}
\hfill \ 
\includegraphics[angle=-90,width=0.49\textwidth,bb=60 20 575 720,clip]{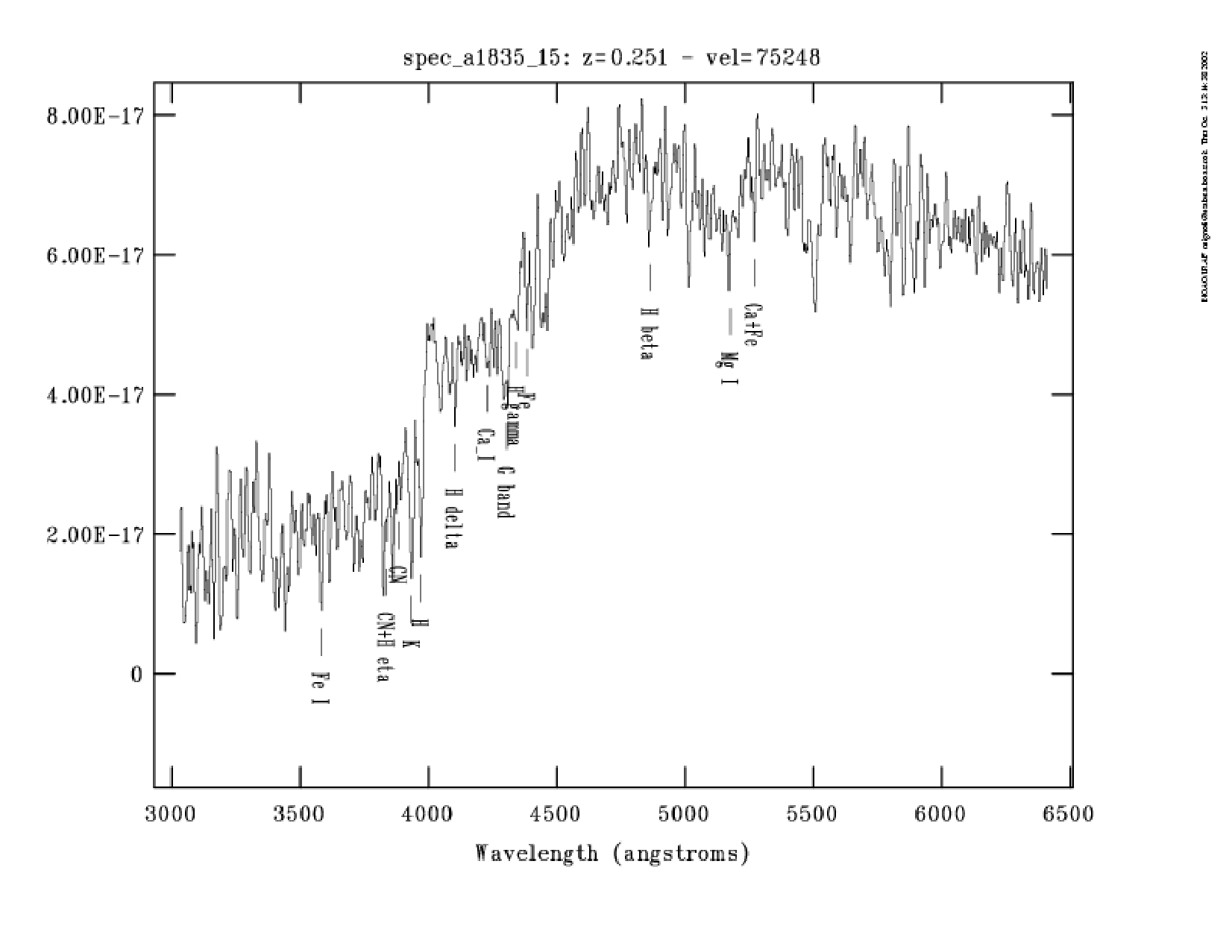}
\caption{Optical spectra of four XBONGs: 
(Top) source \#18 ($z$=0.159, also known as ``P3'', left panel) and \#17 
($z$=0.320, right panel), both in PKS~0312$-$770 field. 
(Bottom) source \#13 ($z$=0.154) in Abell~2690 (left panel) and 
\#15 ($z$=0.250) in Abell~1835 (right panel). 
The principal absorption lines are indicated.}
\label{fig3}
\end{figure}

\begin{figure}
\centering
\includegraphics[angle=0,width=90mm]{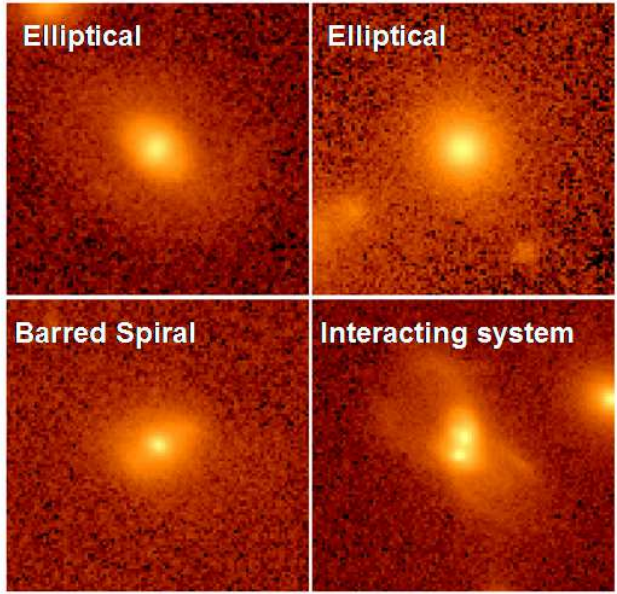}
\caption{$K_{\rm S}$-band images of the four XBONGs shown in Fig.~\ref{fig3} 
obtained with ISAAC at the VLT (seeing$\approx$~0.45\arcsec). 
The two galaxies of the interacting system are separated by 0.9\arcsec, 
corresponding to $\approx$~3.5~kpc.}
\label{fig4}
\end{figure}

Although the XBONGs do not represent a new source population [see, e.g., 
\cite{elv81}], only in the last few years it has become possible to unveil 
a significant number of these sources which were marginally 
represented by a few cases in past studies. 
Since the identification of the first XBONG detected in hard \hbox{X-rays} 
[the source P3 in \cite{fio00} and \cite{com02}; 
see Fig.~\ref{fig3} and \ref{fig4}], many more have been discovered 
(at $z\approx$~0--1) and studied. These sources are characterized by the lack 
of evident AGN signatures in moderate-quality optical 
spectra (see Fig.~\ref{fig3} for four examples from the Hellas2XMM survey), 
although \hbox{X-rays} provide indications for powerful 
\hbox{($L_{2-10~keV}>10^{42}$~\lum)}, not necessarily obscured, AGN emission. 
%
Several interpretations have been proposed to explain the observed properties 
of XBONGs, in particular: 
(a) they are the higher redshift counterparts of local Compton-thick 
galaxies \cite{com03}. This has been suggested by the close 
resemblance of the spectral energy distribution of some XBONGs to that of 
the local Compton-thick galaxy NGC~6240; 
(b) the lack of spectral features at optical wavelengths is due to a 
combination of wide slits, limited-bandpass, low-resolution, and low 
signal-to-noise ratio spectra having the effect of diluting the nuclear 
emission (especially in the case of a low-luminosity AGN) 
by the host galaxy starlight [\cite{mor02, sev03}; see also \cite{mor05}];
(c) the XBONGs are powered by advection-dominated accretion flows (ADAF) 
which are not effective in the production of the emission lines \cite{nar04}; 
(d) XBONGs can possibly be associated with BL Lac objects; this seems to be 
a viable explanation in at least one case \cite{bru03}. 
Whatever the explanation for the lack of spectral signatures, it appears 
difficult to ascribe their absence 
purely to the quality and bandpass of the observations. 
Although the dilution of the nuclear emission by the host galaxy could 
play an important role, it is a fact that the upper limits on typical AGN 
emission lines (i.e., [OIII] and H$\alpha$) measured in the fairly good 
signal-to-noise ratio spectra of the XBONGs sampled by Hellas2XMM (see Fig. 3) are 
an order of magnitude less than the fluxes expected from the observed 
X-ray luminosity. 

Finally, further evidence of the large variety in the properties of 
XBONGs comes from near-infrared (NIR) imaging [see Fig.~\ref{fig4} 
and \cite{mig05}]. 
Despite the optical spectra, which are all typical of early-type galaxies, 
the NIR images show different morphologies: two are ellipticals, one is 
a barred spiral, and one consists of two close interacting galaxies 
(see Fig.~\ref{fig4}.) 
Further broad-band studies and, possibly, a more effective definition 
of ``xbongness'' (e.g., a limit to the equivalent width of any optical 
emission line) are required to understand this ``exotic'' source population 
in more detail. 
%
%

\end{document}